\documentclass[authoryear,preprint,review,12pt]{elsarticle_arxiv}

\usepackage{amssymb,amsmath}
\usepackage{bm}
\usepackage{bbold}
\journal{Icarus}

\begin{document}

\begin{frontmatter}

  \title{The atmospheric impact trajectory of asteroid 2014~AA}

  \author[jpl]{D. Farnocchia}
  \ead{Davide.Farnocchia@jpl.nasa.gov}
  \author[jpl]{S.~R. Chesley}
  \author[ontario]{P.~G. Brown}
  \author[jpl]{P.~W. Chodas}
  \address[jpl]{Jet Propulsion Laboratory, California Institute of
    Technology, Pasadena, CA 91109, USA}
  \address[ontario]{University of Western Ontario, London, Ontario N6A
    3K7, Canada}

\begin{abstract}
  Near-Earth asteroid 2014~AA entered the Earth's atmosphere on 2014 January 2, only 21 hours after being discovered by the Catalina Sky Survey.
  In this paper we compute the trajectory of 2014~AA by combining the available optical astrometry, seven ground-based observations over 69 minutes, and the International Monitoring system detection of the atmospheric impact infrasonic airwaves in a least-squares orbit estimation filter.
  The combination of these two sources of observations result in a tremendous improvement in the orbit uncertainties.
  The impact time is 3:05 UT with a 1$\sigma$ uncertainty of 6 min, while the impact location corresponds to a west longitude of $44.7^\circ$ and a latitude of $13.1^\circ$ with a 1$\sigma$ uncertainty of 140 km.
  The minimum impact energy estimated from the infrasound data and the impact velocity result in an estimated minimum mass of 22.6 t.
   By propagating the trajectory of 2014~AA backwards we find that the only window for finding precovery observations is for the three days before its discovery. 
\end{abstract}


\end{frontmatter}

\section{Introduction}
Small asteroids reach the Earth on a regular basis, collisions with five-meter asteroids, for example, happen on average once every couple years \citep{neopop}.
As of this writing, only two asteroids, both a few meters in size, have been discovered prior to impact: 2008~TC$_{3}$ and 2014~AA.
The first of these was quickly recognized as an impactor, and as a consequence, a large number of observations were collected, which enabled detailed orbit and physical characterization before the atmospheric entry above the Nubian Desert in Sudan \citep{2008tc3}.

On the other hand, for 2014~AA was not timely identified as an impactor and therefore only seven astrometric observations over 69 minutes were collected about 21 hours before the atmospheric impact, which took place on 2014 January 2.\footnote{http://www.nasa.gov/jpl/asteroid/first-2014-asteroid-20140102}
Despite the short observation arc, \citet{sysran} show that using systematic ranging it would have been possible to predict the impact, though with a large uncertainty in the impact location.
The 2014~AA impact was also detected by the infrasound component of the International Monitoring System \citep[IMS][]{ims}. 
As described in Sec.~\ref{s:infrasound}, these infrasound data show that 2014~AA entered the atmosphere over the mid-Atlantic ocean.

In this paper we combine both astrometric and infrasound data to estimate the trajectory and atmospheric impact circumstances of 2014~AA.

\section{Observational data}

\subsection{Optical astrometry}
R. Kowalski \citep{mpec_aa} of the Catalina Sky Survey \citep{catalina} discovered 2014~AA on 2014 January 1 about 21 hours before the asteroid reached the Earth.
The first observation took place at 6:18 UTC.
Since the plane-of-sky rate of motion was large, 4.5$^\circ$/d, the object was likely to be a near-Earth asteroid, and so Kowalski obtained additional observations.
The full observational dataset contains two tracklets, i.e., two sequences of consecutive astrometric observations: the first tracklet contains three observations over 28 minutes, starting 36 minutes later the second tracklet has four additional observations over 5 minutes.
No additional follow-up observations were collected, so the entire astrometric dataset comprises these seven observations\footnote{http://www.minorplanetcenter.net/tmp/2014\_AA.dat} spanning 13$'$ in the plane-of-sky from 6:18 to 7:28 UTC.

\begin{figure}
\centerline{\includegraphics[width=10 cm]{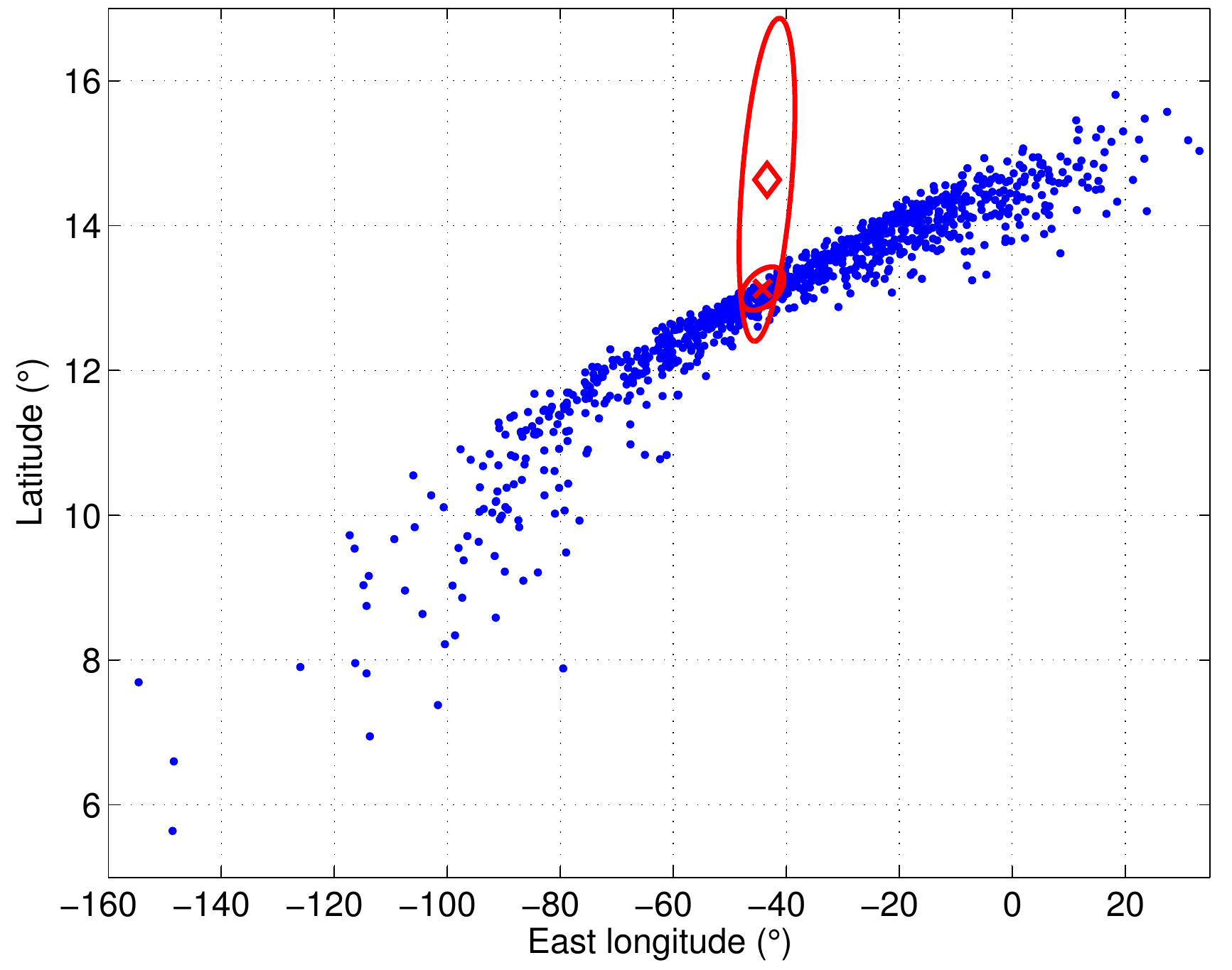}}
\caption{Atmospheric impact location for 2014~AA. Dots correspond to the impact location predictions for an altitude of 50 km obtained from the optical astrometry by using systematic ranging. The diamond marks the nominal infrasound REB measurement surrounded by its 3$\sigma$ uncertainty ellipse. The cross and the surrounding 3$\sigma$ ellipse are for the final estimate, which is presented in Sec.~\ref{s:final}. Note that the axes do not use the same scale.}\label{f:latlon}
\end{figure}

\begin{figure}
\centerline{\includegraphics[width=10 cm]{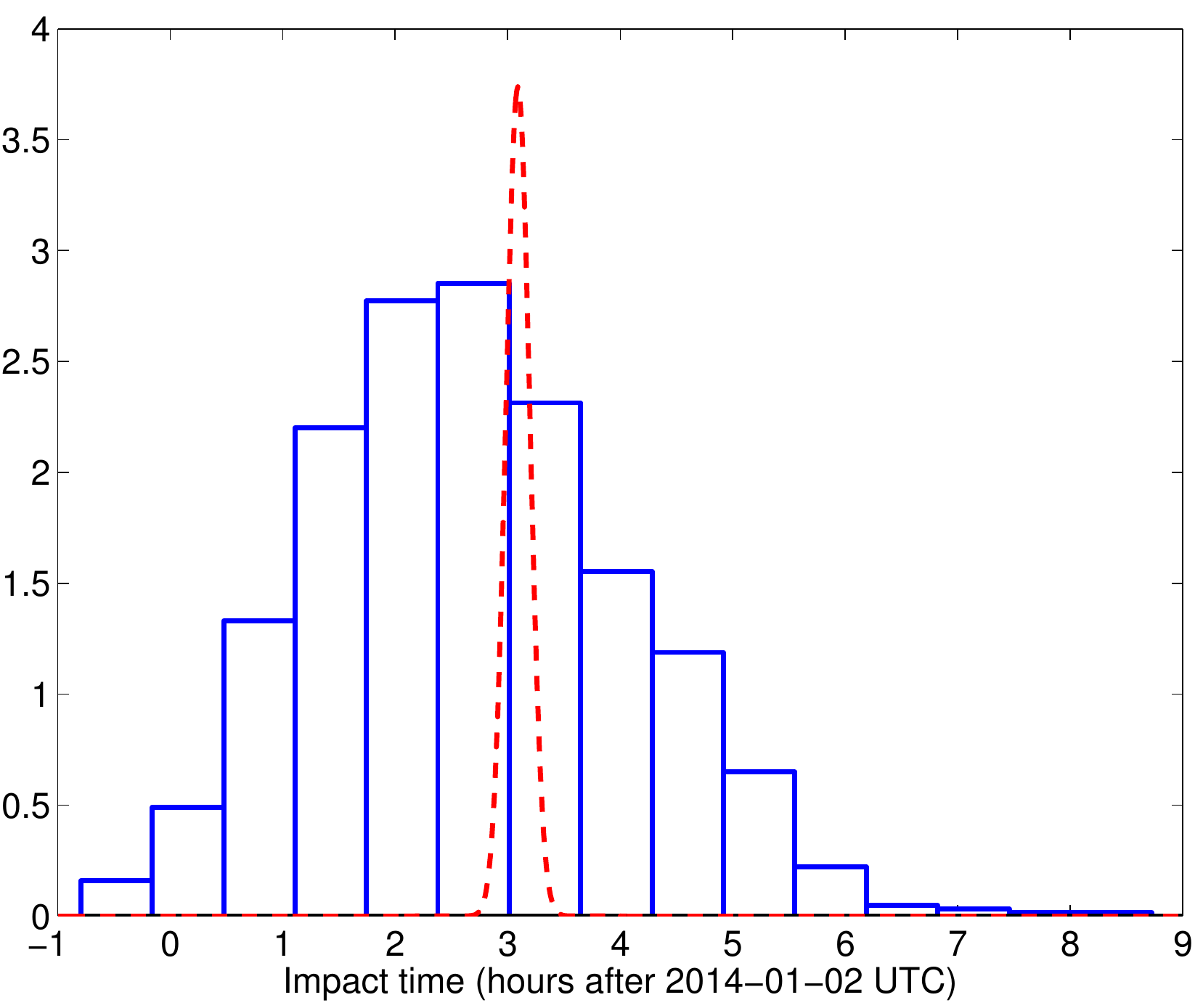}}
\caption{Impact times for 2014~AA. The histograms correspond to the impact times at an altitude of 50 km obtained from the optical astrometry by using systematic ranging. The dashed curve is the probability density function corresponding to the infrasound REB measurement.}\label{f:timp}
\end{figure}

Short observation arcs cause severe degeneracies in the orbit determination process.
Though it is possible to find a least squares solution for 2014~AA from the astrometric data, the uncertainty region cannot be accurately described by an ellipsoid as one would obtain from a linear approximation.
A technique that overcomes these degeneracies is systematic ranging \citep{sysran}, which directly uses the constraint on plane-of-sky motion and rate resulting from the optical observations and scans a raster in topocentric range and range rate to obtain an orbital distribution.
\citet{sysran} apply systematic ranging to 2014~AA and obtain a 100\% impact probability, though the impact location is very uncertain.
Figures~\ref{f:latlon} and \ref{f:timp} show the corresponding distribution of impact location and time at an altitude of 50 km.
These distributions were obtained with 0.5$''$ data weights, which is consistent with the typical quality of the observations obtained at the Mount Lemmon station \citep[][observatory code G96]{deb_pm}.

\subsection{Infrasound data}\label{s:infrasound}
The infrasonic airwaves produced by the 2014~AA atmospheric impact were detected by the infrasound component of the IMS operated by the Comprehensive Nuclear Test Ban Treaty Organization (CTBTO). The IMS is a global system of sensors designed to monitor compliance for the zero-yield CTBTO and includes seismic, infrasound, radionuclide and hydroacoustic technologies \citep{ims}.
Of these technologies, infrasound is a particularly efficient means of detecting atmospheric explosions, including shock waves produced from small asteroid airbursts \citep{edwards10}.
Infrasound is sound below the frequency threshold of normal human hears ($f<20$ Hz) but above the natural buoyancy frequency of the atmospheric ($f \sim 0.001$ Hz).
At such low frequencies, the atmosphere has very low attenuation \citep{beer74} allowing long distance acoustic propagation and hence global coverage for atmospheric explosions using only a few dozen properly spaced stations. 

\begin{figure}
\centerline{\includegraphics[width=11 cm]{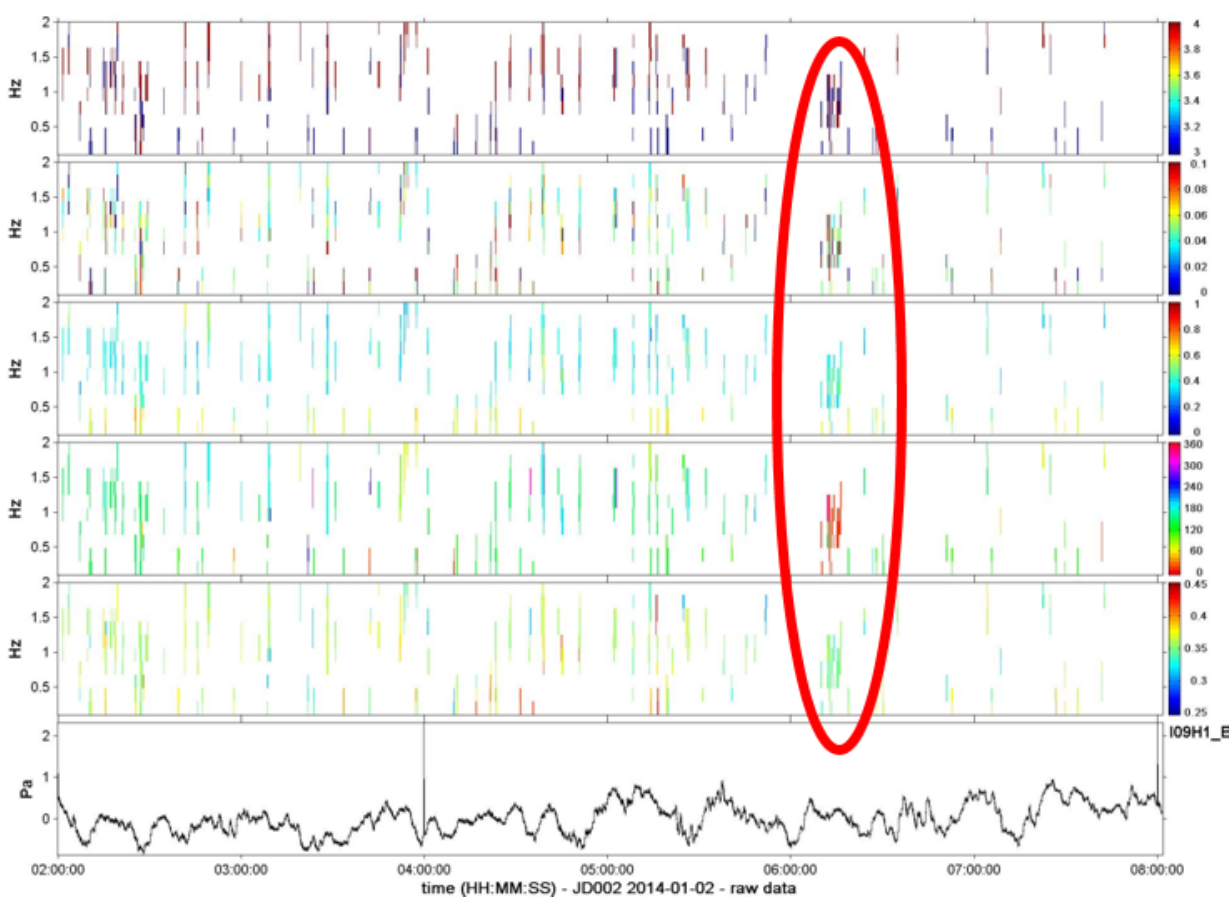}}
\caption{The infrasound signal arrival of the 2014~AA atmospheric impact at I09BR in Brazil (highlights within the red vertical ellipse).
Analysis is performed in time (horizontal) and frequency (vertical) windows producing ``pixels'' of detection.
In this case the time windows are 30~s in length with 90\% overlap and the frequency band is 0.1 -- 2 Hz with 0.2 Hz spacing.
Groupings of similar pixels are identified as a coherent signal arrival \citep[termed a family,][]{brachet}.
The top panel shows the number of sensors in the four element array that participate in the detection -- the higher this value (closer to four) the more likely the detection is real and not a noise fluctuation (note that three is the minimum number required).
The second panel shows signal consistency \citep{runco} -- a coherent signal shows a higher consistency value across the array.
The third panel is the normalized correlation coefficient across the array (higher values again more likely to be associated with true signals).
The 4th panel from the top shows the signal arrival azimuth in each time, frequency ``pixel'' -- groupings in x,y (families) having similar back-azimuths  are another characteristics of real coherent signals. 
The second to last panel show the apparent trace velocity across the array for each pixel -- grouping of nearby pixels having similar trace speeds are a final hallmark of true signals.
The bottom panel shows the unfiltered pressure signal relative to the background atmosphere.}\label{f:brazil}
\end{figure}

\begin{figure}
\centerline{\includegraphics[width=13 cm]{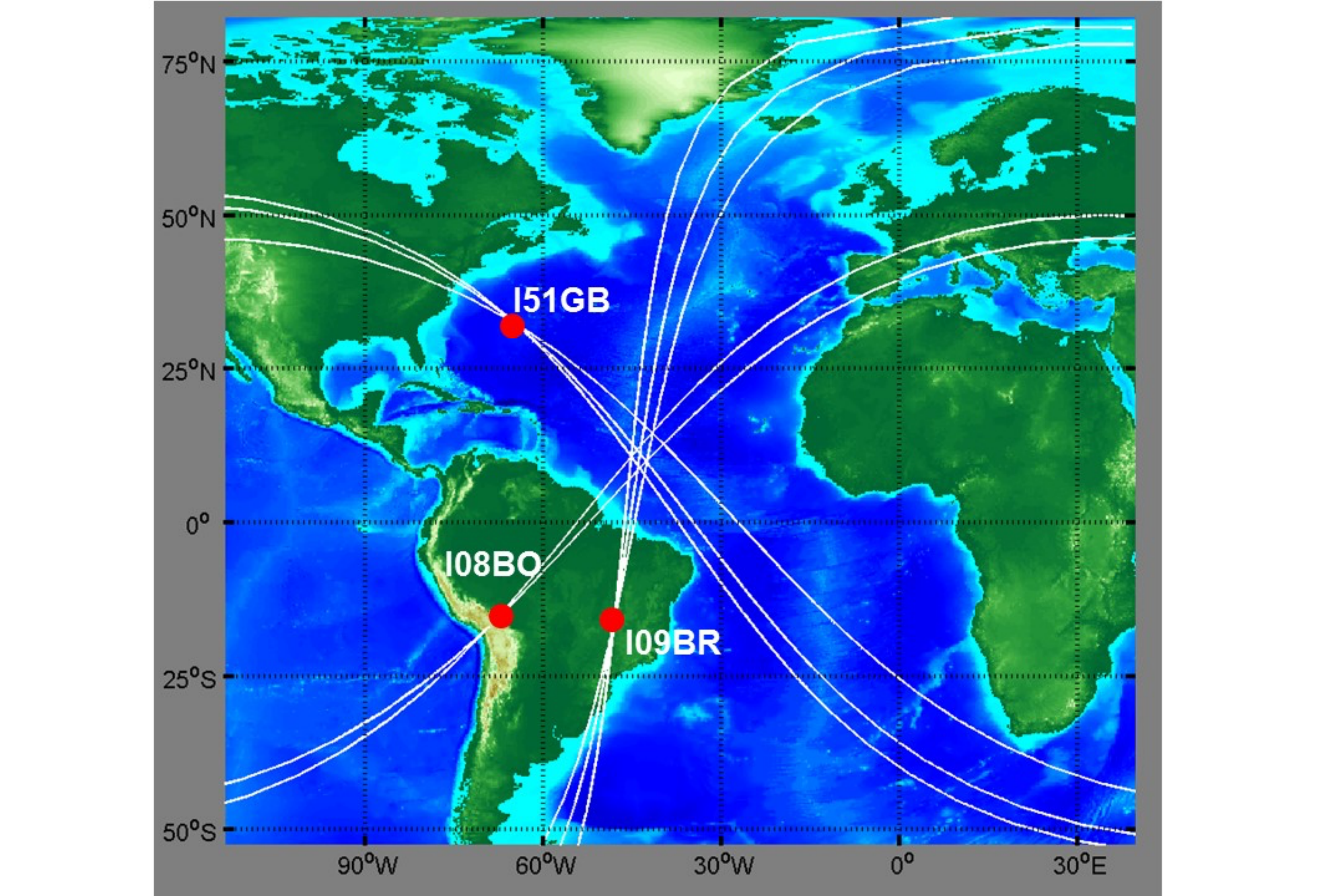}}
\caption{Context map showing location of the three IMS infrasound stations with probable detections of the airwave from 2014~AA. Each great circle line represents the best estimate of the back-azimuth of the signal and its uncertainty as measured manually with PMCC.}\label{f:infrasound}
\end{figure}

At the time of the 2014~AA atmospheric impact, 41 infrasound stations were operational and reporting data to the International Data Centre (IDC). Based on the projected initial ground track of the 2014~AA impact (see Fig.~\ref{f:latlon}), we conducted a manual search of raw infrasound waveform data using the Progressive Multi-Channel Correlation algorithm \citep[PMCC,][]{cansi95} beginning with potential impact points off Africa and moving across the Atlantic. No common signals were detected at the appropriate times from expected azimuths at any stations near Africa. However, as shown in Fig.~\ref{f:brazil}, a likely signal was manually detected with PMCC at I09BR (Brazil) as well as a probable signal at I08BO (Bolivia) and a much weaker signal from Bermuda (I51GB).
The signal back-azimuth and timing were generally consistent with some portions of the projected impact track then available.
The weaker detection at I51GB (Bermuda) was tentatively associated with the event based primarily on good common azimuths. Figure~\ref{f:infrasound} shows the initial geolocation using the direct back-azimuths computed from each array. 

The airwave from 2014~AA was automatically registered by the IMS and published 12 hours after the event as part of the daily Standard Event List level 3 (SEL3) for 2014 January 2.
This automated detection and its associated characteristics are given in Table~\ref{t:sel3}.
\begin{table}[t]
\begin{center}
\begin{tabular}{cccccc}
  \hline
  Time  & Latitude & Longitude & SMA & SMI & Ell. Az\\
  UTC & $^\circ$ N & $^\circ$ W & km &  km & $^\circ$\\
  \hline 
  03:24:31 $\pm$ 4155 s & 11.3047 & 42.9745 & 1699 & 369 & 24\\
  \hline
\end{tabular}
\vspace{0.1cm}

\begin{tabular}{cccccc}
  \hline
  Station  & Range & Arrival & Az$_\text{res}$ & T$_\text{res}$ & SNR\\
   & km & UTC & $^\circ$ & s & \\
  \hline 
   I09BR & 3031 & 06:09:50 & $-$1.0 & $-$74.2 & 2.2\\
   I08BO & 4140 & 07:13:40 & $-$1.3 & 101 & 1.9\\
  \hline
\end{tabular}

\end{center}
\caption{Hypocenter SEL3 location solution for 2014~AA on 2014 January 2 (top) and individual station detections (bottom).
The origin time for the best fit is shown together with the time uncertainty at the 90th percentile level. The location estimate and the semimajor (SMA) and semiminor (SMI) ellipse length is given together with the azimuth orientation of the major axis of the confidence ellipse. The hypocenter azimuth residuals between the observed signal and the best fit solution are shown together with the travel time residuals.
The SNR is the signal to noise ratio at the signal maximum amplitude relative to the noise background outside the signal window.}
\label{t:sel3}
\end{table}
Based on the standard association parameters (such as arrival-quality and probability of detection tests used by the IMS) the I09BR and I08BO signals were correlated to a single event.
The I51GB signal was not included at this stage.
As shown in Table~\ref{t:sel3}, from the station specific estimated signal onset time and azimuthal direction observed at I08BO and I09BR a hypocenter location and origin time were determined using an iterative non-linear least squares inversion \citep{bratt}, which includes the arrival time and back-azimuth.
The propagation speed from a grid of test source locations are tested until a best fit to the observed travel times at each receiver station is found.
These speeds are based on seasonal averages together with spatially dependent raytrace results as described in \citep{brown00}.
Together with the azimuth information, the arrival times are used to find a hypocenter using all stations with a common detection following the basic technique as described in \citet{minster}.
The resulting uncertainties given in Table~\ref{t:sel3} are 90\% confidence bounds.
Residuals in azimuth, and travel-time between the observed and predicted hypocenter are also given in  Table~\ref{t:sel3}.
Note that for these infrasound hypocenter solutions no altitude or depth information is available; at typical infrasound source-station distances a determination of the source height is in general not possible with infrasound only data \citep{silber}.

Several weeks after the SEL3 event was identified, an IDC infrasound analyst reviewed the event, added a probable detection of the event at I51GB and updated the solution as part of the Reviewed Event Bulletin (REB) for 2014 January 2.
As one would expect, the solution based on three stations, shown in Table~\ref{t:reb}, is considerably more constrained, though this is at the expense of larger travel time residuals using standard infrasound propagation speeds. Raytracing using the actual atmosphere might reduce this spread. 
Estimating the source yield from these data is problematic due to the very low SNRs.
However, a crude lower-limit to the source yield can be found by making use of the bolide yield vs range curve developed by \citet[][see their Fig.~12]{ens}.
Based on the positive detection from I08BO at a range of about 4400 km we arrive at a lower limit for the energy of 0.4 kt TNT equivalent. 

\begin{table}[t]
\begin{center}
\begin{tabular}{cccccc}
  \hline
  Time  & Latitude & Longitude & SMA & SMI & Ell. Az\\
  UTC & $^\circ$ N & $^\circ$ W & km &  km & $^\circ$\\
  \hline 
  03:05:25 $\pm$ 632 s & 14.6326 & 43.4194 & 309 & 155 & 76\\
  \hline
\end{tabular}
\vspace{0.1cm}

\begin{tabular}{cccccc}
  \hline
  Station  & Range & Arrival & Az$_\text{res}$ & T$_\text{res}$ & SNR\\
   & km & UTC & $^\circ$ & s & \\
  \hline 
   I09BR & 3386 & 06:09:50 & 1.0 & $-$97.7 & 2.2\\
   I08BO & 4383 & 07:10:30 & 2.7 & 252.7 & 1.9\\
   I51GB & 2919 & 05:44:20 & 5.7 & $-$88.2 & 1.6\\	
  \hline
\end{tabular}
\end{center}
\caption{Hypocenter REB location solution for 2014~AA on 2014 Jan 2 (top) and individual station detections refined with analyst post-processing (bottom).
Columns have the same meaning as given in Table~\ref{t:sel3}.}
\label{t:reb}
\end{table}

Figures~\ref{f:latlon} and \ref{f:timp} show how the REB impact location and time compare to the predictions obtained by systematic ranging applied to the 2014~AA optical astrometry.
The two agree within the uncertainties and the most significant offset is in latitude.

\section{Orbit determination least-squares filter}
Figure~\ref{f:latlon} indicates that if we could combine infrasound and astrometry we would be able to better constrain the orbit of 2014~AA and the impact parameters.
However, while optical astrometry is the most common observable used to determine small body orbits \citep{ast4}, impact observables have only been once used for this purpose, namely the impact times were used in the orbit determination for 16 of the fragments of comet Shoemaker-Levy 9 \citep{SL9}.
%

The standard statistical orbit determination theory \citep[e.g.,][]{bierman, tapley, ast4} relies on the information matrix:
\[
I_{OBS} = \frac{\partial \bm\nu}{\partial \mathbf x}^T W_{OBS} \frac{\partial \bm\nu}{\partial \mathbf x}
\]
where $W_{OBS}$ is the observational weight matrix (i.e., the inverse of the observational covariance), $\bm\nu$ are the observed minus computed (O $-$ C) residuals, and $\mathbf x$ the estimated parameters.
Therefore, $I_{OBS}$ expresses the relationship between the information coming from the observations and the parameters to be estimated.
The least-squares solution for $\mathbf x$, which minimizes the cost function $\bm\nu^T W_{OBS}\bm\nu$, can be found by iteratively solving the normal equation:
\begin{equation}\label{e:normal}
I_{OBS}\Delta\mathbf x = C_{OBS}\ \ , \ \ C_{OBS} = -\frac{\partial \bm\nu}{\partial \mathbf x}^T W_{OBS} \bm\nu\ .
\end{equation}

With independent observations we can sum the contributions to the information matrix, so combining astrometry (AST) and infrasound (REB) yields:
\[
I_{OBS} = \frac{\partial \bm\nu_{AST}}{\partial \mathbf x}^T W_{AST} \frac{\partial \bm\nu_{AST}}{\partial \mathbf x} + 
\frac{\partial \bm\nu_{REB}}{\partial \mathbf x}^T W_{REB} \frac{\partial \bm\nu_{REB}}{\partial \mathbf x}\ .
\]
Similarly, the right-hand-side of Eq.~\eqref{e:normal} is computed as:
\[
- \frac{\partial \bm\nu_{AST}}{\partial \mathbf x}^T W_{AST}\ \bm\nu_{AST} - \frac{\partial \bm\nu_{REB}}{\partial \mathbf x}^T W_{REB}\ \bm\nu_{REB}
\]

For the astrometry we have 14 scalar observations (each optical observations contains two angular measurements) and we simply use a diagonal covariance matrix:
\[
W_{AST} = \frac{1}{(0.5'')^2}\mathbb{1}_{14}
\]
where $\mathbb{1}_{14}$ is the identity matrix of rank 14.

For the infrasound data, we start from the uncertainty information from Table~\ref{t:reb}.
First, we need to convert the 90\% uncertainties to 1$\sigma$: the time uncertainty we have to divide by a factor 1.64, for the the ellipse parameters by a factor 2.15 (two degrees of freedom).
Then, we rotate the ellipse and divide by the Earth radius to obtain the longitude and latitude uncertainty.\footnote{To obtain the longitude uncertainty we need to divide by $\cos(\text{latitude})$ factor}
The corresponding weight matrix is:
\[
W_{REB} = \begin{pmatrix}
	(384\text{ s})^2 & 0 & 0\\
	0 & (1.31^\circ)^2 & 0.33(1.31^\circ)(0.70^\circ)\\
	0 & 0.33(1.31^\circ)(0.70^\circ) & (0.70^\circ)^2\\
	\end{pmatrix}^{-1}.
\]
where the first row is for the time, the second for the longitude, and the third for the latitude.
Note that no observational correlation between impact time and the impact location is available.
Though it is reasonable to expect some observational correlation, treating the impact time as independent of latitude and longitude is conservative as the volume of the uncertainty region is larger.

The parameters to be estimated $\mathbf x$ include the six orbital elements of 2014~AA: perihelion distance $q$, eccentricity $e$, inclination $i$, longitude on node $\Omega$, argument of perihelion $\omega$, and time of perihelion $t_P$.
Moreover, we also include as seventh parameter the atmospheric altitude $h$ corresponding to the infrasound measurement.
The infrasound data do not directly constrain $h$ nor do we expect to find significant signal as a result of the fit.
However, including $h$ in the list of estimated parameters ensures that the uncertainty in the estimated orbital elements accounts for the dependence on $h$ and its uncertainty.

To keep the atmospheric altitude from diverging to non-physical values, we used an a priori value of $h_{APR} = 40$ km and an a priori uncertainty $\sigma_{h_{APR}}$ = 13.33 km.
This a priori constraint gives a 3$\sigma$ interval between 0 and 80 km, which is consistent with the results reported by \citet{brown15}.
To incorporate the a priori information in the least-squares process, Eq.~\eqref{e:normal} becomes:
\[
(I_{OBS} + I_{APR})\Delta\mathbf x = C_{OBS} + C_{APR}
\]
where $I_{APR}$ is a $7\times 7$ matrix whose the only non-zero element is $I_{APR}(7,7) = 1/\sigma_{h_{APR}}^2$, and $C_{APR}$ is a vector of length 7 whose the only non-zero element is $C_{APR}(7) = -(h - h_{APR})/\sigma_{h_{APR}}$.

To complete our least-squares filter we computed the partial derivatives of $\bm\nu_{AST}$ and $\bm\nu_{REB}$ with respect to $\mathbf x$ with a finite difference approximation:
\[
\frac{\partial \bm\nu}{\partial \mathbf x} = \frac{\bm\nu(\mathbf x + \Delta\mathbf x) - \bm\nu(\mathbf x - \Delta\mathbf x)}{2\Delta\mathbf x}\ .
\]
This approach makes it easy to compute the partial derivatives of the infrasound observation: for a given $\mathbf x$ we simply compute the impact time and coordinates and compare to the infrasound measurement. After testing the convergence of the partial derivatives we chose the following steps to compute the finite differences:
\[
(\Delta q, \Delta e, \Delta i, \Delta \Omega, \Delta \omega, \Delta t_p, \Delta h) = (150 \text{ km}, 10^{-5}, 5'', 0.5'', 0.5'', 10 \text{ s}, 10 \text{ km})\ .
\]

\section{Final trajectory estimate}\label{s:final}
Using the best-fit to the optical astrometry obtained from systematic ranging as first guess, we found a converging least squares orbit in five iterations. Table~\ref{t:orb} shows the orbital elements and the corresponding 1$\sigma$ marginal formal uncertainties.
This solution has much smaller formal uncertainties than those obtained only using the astrometry.
For instance, the uncertainty in eccentricity is reduced by a factor of 4, while the uncertainty in the time of perihelion is reduced by a factor of 15.
\begin{table}[t]
\begin{center}
\begin{tabular}{cc}
\hline
 Epoch TDB & 2014-01-01\\
 Eccentricity & 0.21090 $\pm$ 0.00424\\
 Perihelion distance & 0.91659 $\pm$ 0.00156 au\\
 Time of perihelion TDB & 2014-02-15.7198 $\pm$ 0.0182 d\\
 Longitude of node & 101.61301$^\circ$ $\pm$ 0.00967$^\circ$\\
 Argument of perihelion & 52.316$^\circ$ $\pm$ 0.191$^\circ$\\
 Inclination & 1.4109$^\circ$ $\pm$ 0.0292$^\circ$\\
  \hline
\end{tabular}
\end{center}
\caption{Estimated orbital elements of 2014~AA and 1$\sigma$ marginal formal uncertainties.}
\label{t:orb}
\end{table}
The estimated value of the atmospheric impact altitude $h$ is $40.15 \pm 13.33$ km.
The a posteriori uncertainty is the same as the a priori one, thus confirming that there is no signal for $h$ in the observations.
Table~\ref{t:imppar} shows the estimate of the impact parameters and the corresponding uncertainties.
The longitude and latitude estimate is also plotted in Fig.~\ref{f:latlon}.

The converging orbit provides a good match to the observational data (17 scalar observations) with a $\chi^2 = 5.31$, which corresponds to a normalized RMS of $\sqrt{\chi^2/17} = 0.56$. In particular, the RMS of the astrometric residuals is $0.11''$, which is well within the assumed $0.5''$ uncertainties.
For the infrasound data, the time and longitude residuals are 35~s and $0.78^\circ$, both well below the measurement uncertainty.
On the other hand, there is a $\sim$2$\sigma$ discrepancy in latitude with a residual of $1.50^\circ$. As shown from Fig.~\ref{f:latlon} there is a sort of upper bound in the impact location for the solutions obtained from the astrometry.
Crossing this bound would result in unacceptably high residuals for the optical astrometry.
We consider this discrepancy reasonable given the measurement uncertainties.

\begin{table}[t]
\begin{center}
\begin{tabular}{lr}
\hline
\textit{Impact parameters} & \\
 Time UTC & 2014 Jan  2 03:04:50 $\pm$ 373 s\\
 East Longitude & $-44.20^\circ \pm 1.27^\circ$\\
 Latitude & $13.13^\circ \pm 0.10^\circ$\\
 Velocity & $12.070 \pm 0.056$ km/s\\
 Azimuth & $276.67^\circ \pm 0.45^\circ$\\
 Elevation & $77.86^\circ \pm 1.31^\circ$\\
 \hline
 \textit{Impact ellipse} & \\
 1$\sigma$ semimajor axis & 141 km\\
 1$\sigma$ semiminor axis & 10 km\\
 Ellipse Azimuth & 88$^\circ$\\
 \hline
 \textit{Correlation coefficients} & \\
 Time/Longitude & $-0.02$\\
 Time/Latitude & $-0.12$\\
 Longitude/Latitude & 0.44\\
 \hline
\end{tabular}
\end{center}
\caption{Impact parameters for 2014~AA and 1$\sigma$ marginal formal uncertainties. Note that the impact velocity, Azimuth, and Elevation are relative to the body-fixed velocity of the impact location.}
\label{t:imppar}
\end{table}

\begin{figure}
\centerline{\includegraphics[width=10 cm]{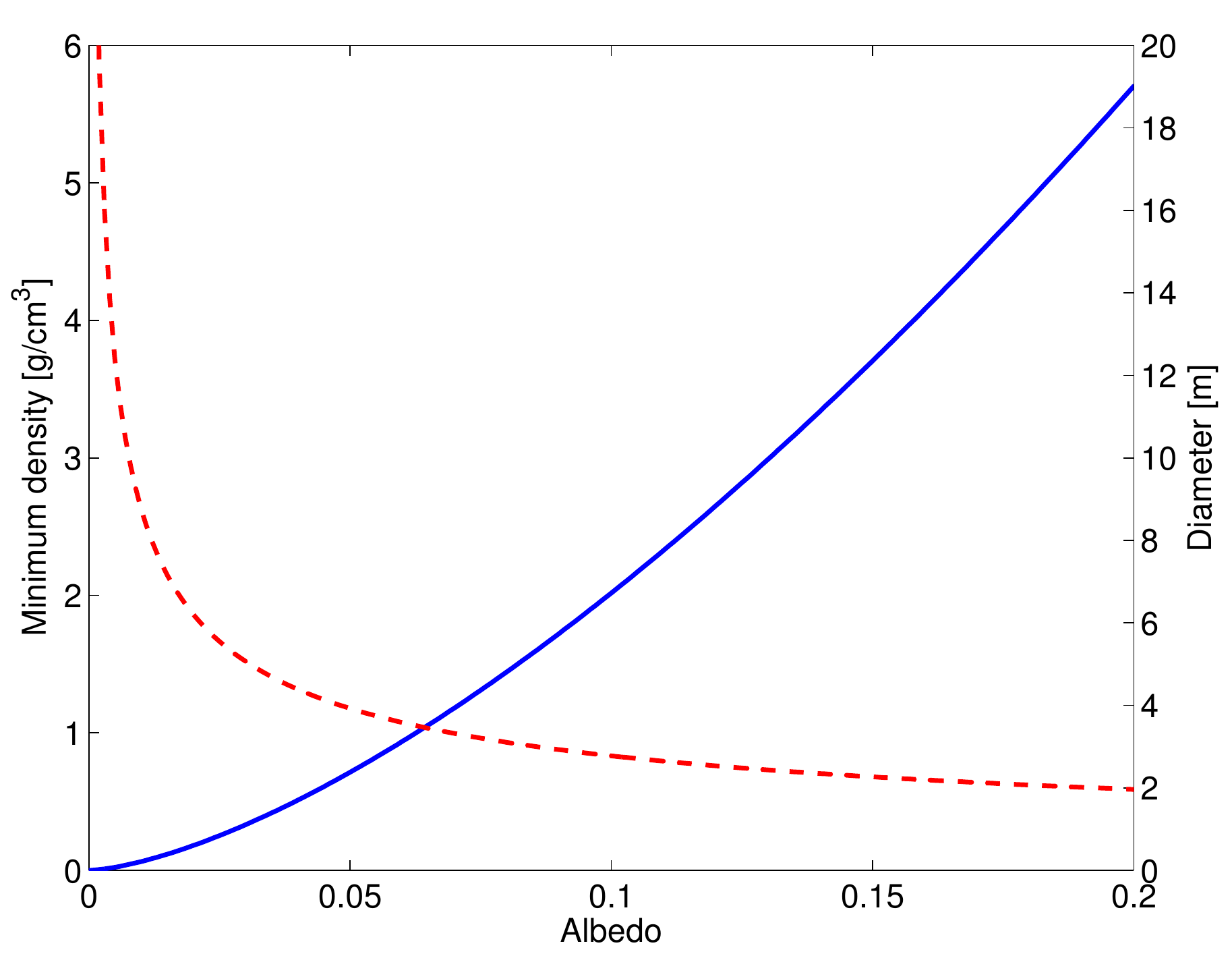}}
\caption{Minimum density (solid line) and diameter (dashed line) of 2014~AA as a function of the albedo based on the 0.4 kt TNT minimum impact energy as estimated from the infrasound data and an absolute magnitude of 30.9.}\label{f:density}
\end{figure}

By combining the lower bound of the impact energy, i.e., 0.4 kt TNT equivalent, and the geocentric impact velocity of 12.17 km/s, we obtain a minimum mass of 22.6 t. Figure~\ref{f:density} shows the corresponding minimum density as a function of the albedo for an absolute magnitude of $H = 30.9$. To obtain reasonable values of the density \citep{carry} the albedo has to be smaller than 20\% and is more likely smaller than 0.15.

\begin{figure}
\centerline{\includegraphics[width=10 cm]{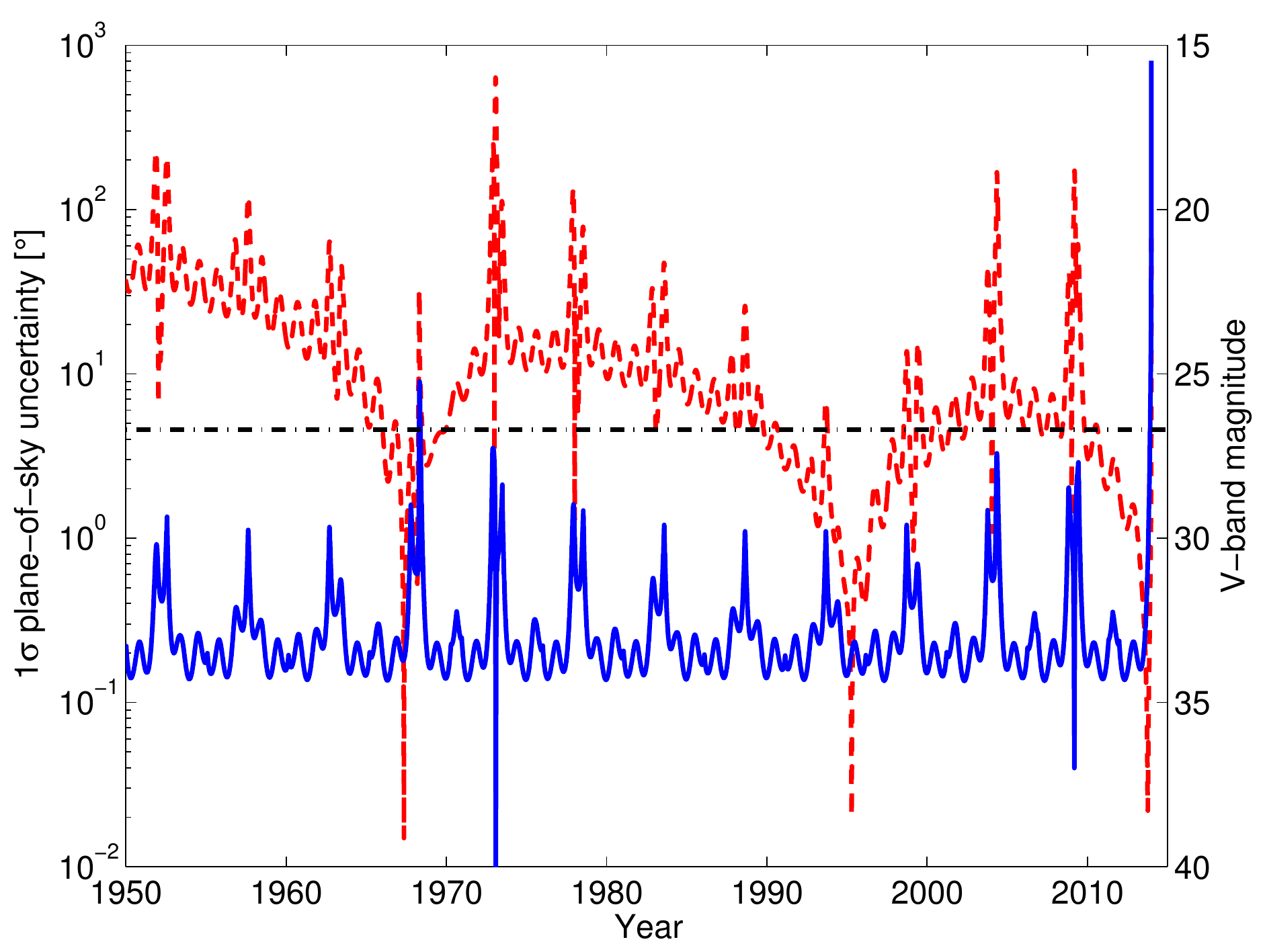}}
\caption{Geocentric V-band (solid line) and semimajor axis of the plane-of-sky 1$\sigma$ uncertainty (dashed) as a function of time. The dash-dotted line corresponds to faintest detection ever performed.}\label{f:eph}
\end{figure}

Figure~\ref{f:eph} shows the evolution of the 2014~AA V-band magnitude and plane-of-sky uncertainty as seen from the geocenter as a function of time.
It is clear why 2014~AA had not been observed before the final approach: the brightest 2014~AA had been since 1950 corresponds to a magnitude $V = 25.2$ in May 1968.
The faintest asteroid observation corresponds to a magnitude of $V = 26.7$ for asteroid 2008~LG$_2$ \citep{micheli}, but this exceptional detection was only possible for targeted follow-up with the telescope tracking at an angular rate matching the asteroid's motion in the sky.
Current and past asteroid search programs do not have the capability of going as fain as $V = 25$ in survey mode, thus ruling out possible observations in past.

During the final approach 2014~AA became brighter than $V=22$ on 2013 December 29, about three days before the discovery observations.
Thus, there is the possibility of extending the arc of the final apparition, though searches in the Pan-STARRS and Catalina archives have not been successful (Wainscoat and Christensen, private communication).

\section{Conclusions}
In this paper we combined ground-based optical astrometry and infrasound data to compute the trajectory of 2014~AA.
It is the first time that these two kinds of observations are combined together to compute the orbit of an Earth-impacting asteroid.
A similar technique could prove useful in the future to reconstruct the orbits of impacting asteroids that are discovered shortly before an impact and thus could have a limited observational dataset.

The orbit of 2014~AA we estimated in this paper has significantly lower uncertainties than that computed only from the astrometric observations. For instance, as shown in Fig.~\ref{f:latlon}, the impact locations derived from the astrometry cover more than 180$^\circ$ in longitude, while the 3$\sigma$ longitude uncertainty obtained combining infrasound and astrometry is less than 4$^\circ$.
The 90\% confidence region of the impact location as estimated from the infrasound observations covers 150\,000 km$^2$, while including the information from the astrometry reduces the extent of this area to 20\,400 km$^2$.
The atmospheric impact altitude remains essentially unconstrained other than a priori constraints based on the analysis of past impact airburst observations \citep{brown15}.

The geocentric impact velocity was 12.17 km/s, which, together with the minimum impact energy of 0.4 kt TNT estimated from the infrasound data, places a lower bound of 22.6 t on the mass of 2014~AA and suggests an albedo smaller than 20\%.

A backward trajectory propagation shows that 2014~AA had never been bright enough to be observable in the last 50 years before the final approach, thus limiting the possibility of finding earlier precovery observations.

\section*{Acknowledgments}
We thank J.~D. Giorgini for useful comments that helped improve the paper.
D. Farnocchia, S.~R., Chesley, and P.~W. Chodas conducted this research at the Jet Propulsion Laboratory,
California Institute of Technology, under a contract with NASA. 
P.~G. Brown received support for this study from NASA Cooperative agreement NNX11AB76A. P.~G. Brown thanks the Canadian Hazard Information Service of Natural Resources Canada for technical support and IDC access as part of the Canadian National Data Center.

\copyright 2016. All rights reserved.

\end{document}